\documentclass[pra,10pt,twocolumn,showpacs,superscriptaddress]{revtex4}
\usepackage{amsmath}
\usepackage{latexsym}
\usepackage{amssymb}
\usepackage[colorlinks=true, citecolor=blue, urlcolor=blue ]{hyperref}
\usepackage{float}
\usepackage{graphicx}
\usepackage{graphics,graphicx,epsfig}
\usepackage{epstopdf}

\newcommand{\be}{\begin{eqnarray}}
\newcommand{\ee}{\end{eqnarray}}

\usepackage{amsfonts}
\usepackage{appendix}
\begin{document}

\title{Structure of quantum and broadcasting nonlocal correlations}

\author{Debashis Saha}
\affiliation{Institute of Theoretical Physics and Astrophysics, University of Gda\'{n}sk, 80-952 Gda\'{n}sk, Poland}

\author{Marcin Paw\l{}owski}
\affiliation{Institute of Theoretical Physics and Astrophysics, University of Gda\'{n}sk, 80-952 Gda\'{n}sk, Poland}


\begin{abstract}
The multipartite setting offers much more complexity of nonlocality than the bipartite one. We analyze the structure of tripartite nonlocal
correlations by proposing inequalities satisfied by each of type: bilocal, broadcasting and quantum, but violated by the other two.
One of the inequalities satisfied by broadcasting correlations is generalized for multipartite systems. The study of its quantum mechanical violation reveals that Greenberger-Horne-Zeilinger-like states exhibit new, powerful correlations. 
\end{abstract}


\maketitle

\section{Introduction} After the formulation of Bell's theorem showing the incompatibility between quantum mechanics and local realism \cite{Bell,Bell1} and the discovery of the famous CHSH inequality \cite{CHSH}, studies of nonlocality naturally evolved to more general scenarios. One direction was to devise inequalities involving more parties \cite{Mermin}; the other to consider stronger than quantum correlations [quantum mechanical (QM)] \cite{PR}. Of particular interest is the intersection of these approaches: generalized multipartite correlations. Research in this area was pioneered by Svetlichny, who introduced the notion of {\it bilocal} (BL) inequalities \cite{s,s1,c}. They are satisfied by every theory in which the set of parties can be divided into two groups sharing only classical correlations, while any type of probability distribution is allowed inside both groups. This "anything goes" behavior in the groups clearly allows for some probability distributions impossible in quantum mechanics, but the most interesting result in \cite{s} is that there are quantum correlations which violate BL inequalities. This leads to the introduction of the notions of genuine multiparty entanglement and nonlocality. The idea of BL correlations was further developed in \cite{acin,gisin,gisin1}, where two, weaker versions of them were introduced: {\it no-signalling bilocal} (NSBL) and {\it time-ordered bilocal} (TOBL) to understand genuine tripartite nonlocality from a better physical as well as operational point of view. Further study of TOBL models lead to a powerful result in the foundations of quantum theory which states that it cannot be derived with only two-partite informational principles \cite{multi,multi2}.

Recently, another category of correlations, namely {\it broadcasting correlations} (BCs) has been proposed \cite{jones, bancal}, in which some of the parties communicate their measurements information to all other. As pointed out by Bancal {\it et al.}, quantum mechanical (QM)  violation of broadcasting correlations serves as a different measure of multipartite nonlocality, and equivalently can also be regarded as an alternative notion of genuine multipartite nonlocality. However, these correlations remained largely unexplored. The aim of this work is to change this state.

Firstly, we demonstrate that for each of the three sets- bilocal, broadcasting and quantum, there are probability distributions which are members of this set but not of the other two. Moreover, we also prove that for each of them there are inequalities which are violated by the other two. This reveals the complicated and interesting structure of tripartite nonlocality. The broadcasting inequality which is violated by both bilocal and quantum correlations, is generalized for the multipartite case and the QM violation is studied.
Our results imply that all the Greenberger-Horne-Zeilinger (GHZ)-like states for $N=3,4,5,6$-partite system, violate broadcasting inequality. We also provide the analogous notion of anonymous quantum nonlocality \cite{aqn}, with respect to broadcasting correlations.

\section{ Different multipartite correlations } We begin by defining different sets of probability distributions. Let us consider a three-partite system where the observables $X,Y,Z$ are measured by the first, second, and third parties yielding outcomes $a,b,$ and $c$ respectively. As mentioned in the previous section, the BL and TOBL correlations are defined as follows \cite{s,gisin,acin}:
\begin{equation}
\begin{split}\label{tobl}
\text{TOBL: ~~} & P(a,b,c|X,Y,Z) = \\& \sum_{\lambda_1} q_{\lambda_1} P_{\lambda_1}(a|X) P^{Y\rightarrow Z}_{\lambda_1}(b,c|Y,Z) \\& +  \sum_{\lambda_2} q_{\lambda_2} P_{\lambda_2}(a|X) P^{Z\rightarrow Y}_{\lambda_2}(b,c|Y,Z) \\& + \sum_{\lambda_3} q_{\lambda_3}
P_{\lambda_3}(b|Y) P^{X\rightarrow Z}_{\lambda_3}(a,c|X,Z)\\& + \sum_{\lambda_4} q_{\lambda_4} P_{\lambda_4}(b|Y) P^{Z\rightarrow
X}_{\lambda_4}(a,c|X,Z)\\&+ \sum_{\lambda_5} q_{\lambda_5} P_{\lambda_5}(c|Z) P^{X\rightarrow Y}_{\lambda_5}(a,b|X,Y) \\& +
\sum_{\lambda_6} q_{\lambda_6} P_{\lambda_6}(c|Z) P^{Y\rightarrow X}_{\lambda_6}(a,b|X,Y)
\end{split}
\end{equation}                                                                                                                           \begin{equation}\label{bl}                                                                                                                                                     \begin{split}                                                                                                                                                                  \text{BL: ~~}& P(a,b,c|X,Y,Z) = \\& \sum_{\lambda_1} q_{\lambda_1} P_{\lambda_1}(a|X) P_{\lambda_1}(b,c|Y,Z) \\& + \sum_{\lambda_2}
 q_{\lambda_2} P_{\lambda_2}(b|Y) P_{\lambda_2}(a,c|X,Z) \\& + \sum_{\lambda_3} q_{\lambda_3} P_{\lambda_3}(c|Z) P_{\lambda_3}(a,b|X,Y). \\
 \end{split}
 \end{equation}                                                                                                                                                         Here $\sum_{\lambda_i} q_{\lambda_i} =1$, $P_{\lambda_1} (b,c|Y,Z)$ is an arbitrary probability distribution, and
$P_{\lambda_1}^{Y\rightarrow Z}(b,c|Y,Z)$ is any one-way signaling distribution where information is sent from second party to third party.
NSBL is the particular case of bilocality where all the joint probabilities, like $P_{\lambda_1} (b,c|Y,Z)$  are nonsignaling. It was shown that
NSBL $\subsetneq$ TOBL $\subsetneq$ BL \cite{gisin,acin}.

Our main focus in this work is the broadcasting correlations in which: $(i)$ all the parties have a local description, and
$(ii)$ in each run of the experiment some of the parties broadcast information to all the others. When there is more than one broadcasting party, a subtlety might arise in this definition, which was not pointed out earlier. In this case, the outcome statistics of two broadcasting parties depend on one another and this might lead to grandfather-like paradoxes or a not well defined probability distribution. Thus we choose to assume a preferred frame of reference.
It is known that in a world with a preferred frame of reference, the possibility of superluminal signaling does not lead to paradoxes, and all events can be ordered chronologically. With respect to this time ordering of the measurement events, we allow some particles to send the information about the choice and the outcome of the measurement to all other parties. Note that we do not know what this preferred frame of reference is, and it can change over time. Moreover, the particles can have access to shared randomness so in each round the time ordering of measurement events may differ. It is easy to see that if more than one particle is broadcasting in the tripartite case, then
every probability distribution is allowed. Therefore, here we concentrate on the case when only one particle, that which is measured first is sending information.

As a consequence, the joint probability $P(a,b,c|X,Y,Z)$ in the tripartite {\it first event broadcasting} scenario, can be written as
\begin{equation}\label{BC3}
\begin{split}
& P(a,b,c|X,Y,Z)\\& = \sum_{\lambda_1} q_{\lambda_1}  P_{\lambda_1}(a|X) P^{X \rightarrow Y}_{\lambda_1}(b|Y,X,a) P^{X \rightarrow Z}_{\lambda_1}(c|Z,X,a) \\& + \sum_{\lambda_2} q_{\lambda_2} P_{\lambda_2}(b|Y) P^{Y \rightarrow X}_{\lambda_2}(a|X,Y,b) P^{Y \rightarrow Z}_{\lambda_2}(c|Z,Y,b) \\& +\sum_{\lambda_3} q_{\lambda_3} P_{\lambda_3}(c|Z) P^{Z \rightarrow Y}_{\lambda_3}(b|Y,Z,c) P^{Z \rightarrow X}_{\lambda_3}(a|X,Z,c).
\end{split}
\end{equation} It is noteworthy that we remain consistent with Bell's original assumption of local realism, where the measurement outcome can be stochastic, while determinism is a derived concept \cite{Bell,Bell1}.

Before we explicate the relation between different tripartite correlations, it is noted that broadcasting (\ref{BC3}) allows two one-way signaling channels whereas TOBL  (\ref{tobl}) allows only one. Hence, TOBL and NSBL are proper subsets of broadcasting correlations.

\section{ Relation between different sets of correlations} Let us consider broadcasting correlations, in which one of the two observables $X,Y,Z \in \{0,1\}$ producing a binary outcome $a,b,c \in \{+,-\}$ is measured on particle 1,2, or 3 respectively. Given the fact that observable the $X$ is measured first on particle 1 and the outcome is $a$, from the definition of broadcasting correlations (\ref{BC3}) one can infer the existence of a joint probability distribution of the measurement outcomes on the other two particles in the form $P_{\lambda_1}(b^{0}, b^{1}, c^{0},c^{1}|a^X)$ related to a particular hidden state $\lambda_1$. The index of lambda being '1' denotes that the first measurement is performed on particle '1'; $b^{0(1)},c^{0(1)}$ denotes the outcomes when $Y=0(1),Z=0(1)$ are measured on particles 2 and 3 respectively. Clearly, the probability of obtaining the outcomes $b,c$ depends on the measurement setting $X$ and the outcome $a$ of the first particle. Similarly, we can define $P_{{\lambda_2}} (a^{0},a^{1}, c^{0},c^{1}|b^Y)$ and $ P_{{\lambda_3}} (a^{0},a^{1},b^{0}, b^{1}|c^Z)$ when particle 2 or 3 is measured first. The observed joint probability of finding, say $a^{0}=+,b^{0}=+,c^{0}=+$ denoted as $ P(a^0+,b^0+,c^0+)$ is given by the expression
\begin{equation}
\begin{split}\label{BC1}
& P(a^0+,b^0+,c^0+)=  \sum q_{\lambda_1} P_{\lambda_1}(a^0+,b^0+,c^0+) \\ & + \sum q_{{\lambda_2}} P_{{\lambda_2}}(a^0+,b^0+,c^0+) +\sum q_{{\lambda_3}} P_{{\lambda_3}}(a^0+,b^0+,c^0+)
\end{split}
\end{equation} for $\sum_{\lambda_1} q_{\lambda_1} + \sum_{\lambda_2} q_{\lambda_2} + \sum_{\lambda_3} q_{\lambda_3}=1$, where  the joint probabilities $P_{\lambda_1}(a^0+,b^0+,c^0+), P_{{\lambda_2}}(a^0+,b^0+,c^0+),$ and $ P_{{\lambda_3}}(a^0+,b^0+,c^0+)$ are marginals of the overall joint probability,
\begin{equation} \label{BC2}
\begin{split}
P_{\lambda_1}(a^0+,b^0+,c^0+) =  \sum_{b^1,c^1 = \pm} P_{\lambda_1} (+,b^1,+,c^1|a^0+) P_{\lambda_1}(a^0+)\\
P_{\lambda_2}(a^0+,b^0+,c^0+) =  \sum_{a^1,c^1 = \pm} P_{{\lambda_2}} (+,a^1,+,c^1|b^0+) P_{\lambda_2}(b^0+) \\
P_{{\lambda_3}}(a^0+,b^0+,c^0+) =  \sum_{a^1,b^1 = \pm} P_{{\lambda_3}} (+,a^1,+,b^1|c^0+) P_{\lambda_1}(c^0+).
\end{split}
\end{equation} In the following, we use the description of broadcasting correlations given by (\ref{BC1})-(\ref{BC2}).\\

\subsection{ Broadcasting inequalities} In order to learn about the structure of tripartite correlations, we now introduce several new inequalities.

{\bf Theorem 1:} {\it In the tripartite scenario described above, the following inequality holds,
 \begin{equation}\label{s1}
\begin{split}
& I = S_3 + S' \overset{BL}{\leq}  5 \overset{BC}{\leq} 6
\end{split}
\end{equation} where $S_3 =  \langle a^0b^0c^0 \rangle + \langle a^0b^0c^1 \rangle +\langle a^0b^1c^0 \rangle - \langle a^0b^1c^1 \rangle + \langle a^1b^0c^0 \rangle - \langle a^1b^0c^1 \rangle - \langle a^1b^1c^0 \rangle - \langle a^1b^1c^1 \rangle$, \cite{cf} and $S' = P(a^0+,b^0+,c^0+) +  P(a^0-,b^0+,c^0+) + P(a^1+,b^0-,c^0-) + P(a^1-,b^0-,c^0-)$.  Moreover, quantum correlations allow for values of $I$ higher than 6}.\\
{\it Proof: } It was shown that, the quantity $S_3$ introduced by Svetlichny \cite{s} as BL inequality, is also bounded by broadcasting correlations \cite{bancal},
\begin{equation}\label{s}
\begin{split}
S_3 \overset{BL/BC}{\leq}  4.
\end{split}
\end{equation} From Eq.(\ref{s}) and the fact that the algebraic maximum value of $S'$ is 2, one concludes the upper bound of (\ref{s1}) for BL and BC cannot be greater than 6. For BL correlations, we checked that $S' \leq 1$ by considering every possible deterministic BL strategy. The general ones are just linear combinations of these. Further, we notice that, for the broadcasting correlation, $\forall \lambda_1$
\begin{equation}\begin{split}
&P_{\lambda_1}(a^0+)=P_{\lambda_1}(a^1-)=1, \\
&P_{\lambda_1}(+,+,+,+|a^0+)=P_{\lambda_1}(-,-,-,-|a^1-) =1,
\end{split}\end{equation} and for all $\lambda_2$ and $\lambda_3$ $q_{\lambda_2}=q_{\lambda_3}=0$, reproduces the upper bound 6.

A simple calculation for the state and measurements for which the maximum violation of (\ref{s}), is obtained \cite{s}, yields $I_{QM} = 4\sqrt{2} + 0.5 \approx 6.157$. QED.

It is also trivial to check that $S'\leq 1$ in any no-signalling theory and hence in QM. Therefore we obtain
\be \label{sprime}
S' \overset{BL/QM}{\leq} 1 \overset{BC}{\leq}  2.
\ee
This shows that there are broadcasting correlations which are neither in BL nor in QM.

\begin{figure}[hbp]
\centering
\includegraphics[height=7cm,width=8cm]{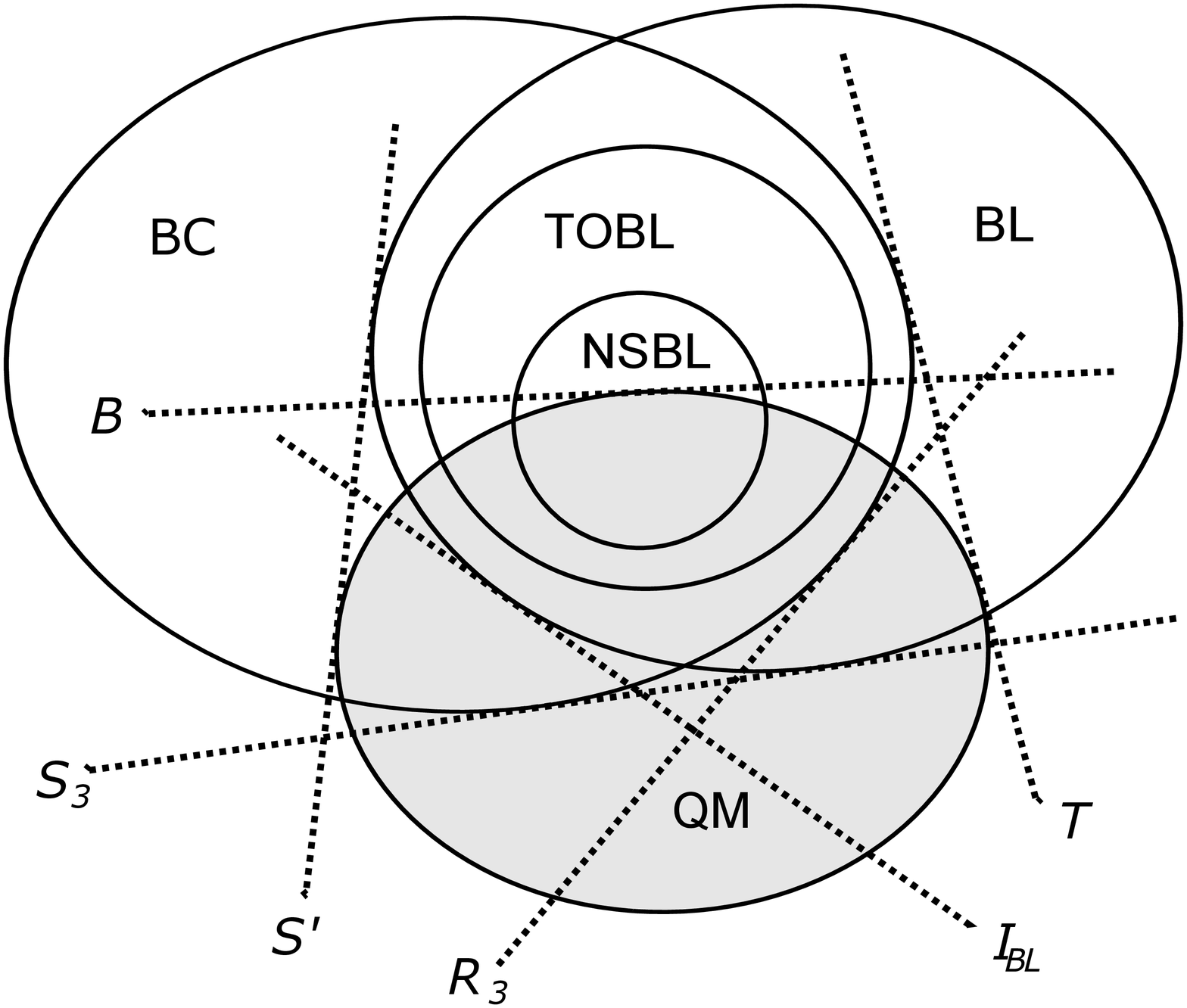}
\caption{Representation of the overall structure of different bilocal, broadcasting and quantum correlations. Dashed lines represent inequalities given by (\ref{s1}),(\ref{s}),(\ref{sprime}),(\ref{r}),(\ref{t}). Here $B$ represents the Bell-CHSH term ($\langle a^0b^0\rangle + \langle a^0b^1 \rangle +\langle a^1b^0 \rangle - \langle a^1b^1 \rangle$) \cite{CHSH}, which is bounded by $2\sqrt{2}$ in QM.}
\end{figure}

{\bf Theorem 2:} {\it In the tripartite scenario described above, broadcasting correlations satisfy the following inequality }
\begin{equation}\label{r}
\begin{split}
&R_3= P(a^0+,b^0+,c^0+) - P(a^1+,b^0+,c^0+) \\&- P(a^0+,b^1+,c^0+) - P(a^0+,b^0+,c^1+) - P(a^0+,b^1-,c^1-) \\& - P(a^1-,b^0+,c^1-) -P(a^1-,b^1-,c^0+)  \leq 0.
\end{split}
\end{equation} {\it Proof: } Since the left-hand side of (\ref{r}) is symmetric under the permutation of parties, it is sufficient to show that (\ref{r}) is satisfied in the context where particle 1 is measured before the others.  If we expand the joint probabilities appearing in (\ref{r}) by using (\ref{BC2}), we obtain
\begin{equation}
\begin{split}
&P_{\lambda_1}(a^1+,b^0+,c^0+)+P_{\lambda_1}(a^0+,b^1+,c^0+) \\& +P_{\lambda_1}(a^0+,b^0+,c^1+) + P_{\lambda_1}(a^0+,b^1-,c^1-)\\&+ P_{\lambda_1}(a^1-,b^0+,c^1-) + P_{\lambda_1}(a^1-,b^1-,c^0+) \\ &
=  P_{\lambda_1}(a^0+) \Big[ \sum_{b^0,c^1 = \pm} P_{\lambda_1} (b^0,+,+,c^1|a^0+) \\& + \sum_{b^1,c^0 = \pm} P_{\lambda_1} (+,b^1,c^0,+|a^0+) \\& + \sum_{b^0,c^0 = \pm} P_{\lambda_1} (b^0,-,c^0,-|a^0+) \Big] +P_{\lambda_1}(a^1+,b^0+,c^0+) \\&
+ P_{\lambda_1}(a^1-,b^0+,c^1-) + P_{\lambda_1}(a^1-,b^1-,c^0+) \\&
= P_{\lambda_1}(a^0+,b^0+,c^0+)+ \mbox{~non-negative~terms} \\&
\geq P_{\lambda_1}(a^0+,b^0+,c^0+).
\end{split}
\end{equation} This equation is true for probabilities indexed with ${\lambda_2}, {\lambda_3}$ due to symmetry. This concludes the proof. QED.

By making projective measurements on the GHZ state, quantum mechanics lets us violate (\ref{r}) and obtain a value of $0.0364$ (more on quantum violations of broadcasting inequalities later). Note that, for the BL correlations (\ref{bl}) in which $\forall \lambda_1,$
\begin{equation}\begin{split}
& P_{\lambda_1}(a^0+)=P_{\lambda_1}(a^1-)=1, \\
&P_{\lambda_1}(b^0+,c^0+)=P_{\lambda_1}(b^1-,c^0-)=P_{\lambda_1}(b^0-,c^1-)\\& =P_{\lambda_1}(b^1+,c^1+) =1
\end{split}\end{equation}and $q_{\lambda_2} = q_{\lambda_3} = 0, \forall \lambda_2, \lambda_3$, the left-hand side of (\ref{r}) is 1.

Obviously, correlations which involve one-way signaling between two parties are in both BC and BL but not in QM. It can also be checked by considering deterministic strategies of one-way signaling between particle '1' to '2' that the following inequality holds,
\begin{equation}\begin{split}\label{t}
&T = P(a^0+,b^0+)+P(a^0-,b^1+) \\& +P(a^1+,b^0-)+P(a^1-,b^1-) \leq 2.
\end{split}\end{equation} Since $T$ contains joint probability of two particles $T \overset{BC/QM}{\leq} 2 $, but the value of $T$ is 4 for BL.
These results give us all the information we need to present a complete relation among broadcasting, different bilocal and quantum correlations, which we do in Fig.1.

\subsection{Generalization to $N$ parties }
We now show that a generalization of (\ref{r}) holds for multipartite system, in which the particle that is measured first simply broadcasts to all the other $N-1$ local particles.
Suppose one of two dichotomic observables is measured on each of the spatially separated particles which are denoted by $X_i$ and yield outcomes $a_i$, where the index $i$ represents the $i$-th particle. Just as in the tripartite case, we can define a joint probability distribution $P_{\lambda_1}(a^{0}_2,a^{1}_2,...,a^{0}_N,a^{1}_N|a^{X_1}_1)$ of the measurement outcomes on all the other particles,  conditioned on the event that $X_1$ is measured first on particle '1' and outcome $a_1$ is observed. A similar probability distribution can be defined for probabilities indexed by $\lambda_2,...,\lambda_N$. We can write the marginal probability distribution $P_{\lambda_1}(a^0_1+,a^0_2+,a^0_3+,...,a^0_N+)$ in terms of the joint one,
\begin{equation}\label{eq9}
\begin{split}
& P_{\lambda_1}(a^0_1+,a^0_2+,a^0_3+,...,a^0_N+) =\\ &
 \sum_{a^1_2,...,a^1_N=\pm} P_{\lambda_1}(+,a^1_2,+,a^1_3,...,+,a^1_N|a^0+),
\end{split}
\end{equation} and the following relation can be observed
\begin{equation}
\label{eq10}
\begin{split}
& P_{\lambda_1}(a^1_1+,a^0_2+,...,a^0_N+)+ P_{\lambda_1}(a^0_1+,a^1_2+,...,a^0_N+)+  \\
& ...+P_{\lambda_1}(a^0_1+,a^0_2+,...,a^1_N+) + P_{\lambda_1}(a^0_1+,a^1_2-,...,a^1_N-) \\
& + P_{\lambda_1}(a^1_1-,a^0_2+,...,a^1_N-)+ ... + P_{\lambda_1}(a^1_1-,a^1_2-,...,a^0_N+) \\
& = P_{\lambda_1}(a^0_1+,a^0_2+,...,a^0_N+) +  \mbox{non-negative terms}
\end{split}
\end{equation} Again, (\ref{eq10}) holds for any index $\lambda_1,...,\lambda_N$ as it is symmetric under the permutation of particles. Thus we obtain the inequality
\begin{equation}\label{rn}
\begin{split}
& R_N = P(a^0_1+,a^0_2+,...,a^0_N+) - \sum^N_{ j=1} P(a^1_j+,a^0_*+)\\
& - \sum^N_{ j=1} P(a^0_j+,a^1_*-) \leq 0
\end{split}
\end{equation} where $a_*$ denotes the joint probability of all particles except the $j$-th one. \\

\section{Quantum mechanical violation} To show the QM violation of the inequality given by (\ref{rn}), we take the family of GHZ-like states
\begin{equation}\label{ghz}|GHZ\rangle = \cos(t)|0\rangle^{\otimes N} + \sin(t) |1\rangle^{\otimes N}
\end{equation} shared by spatially separated parties \cite{GHZ}, where $t \in (0,\frac{\pi}{4}]$. We consider the measurement settings on each particle to be projective and the angle between the measurements corresponding to different settings of one party is the same for all parties. Moreover, the angle with the $z$-direction is the same for all parties. Therefore we can parametrize them by angles $\phi_i, \alpha, \beta, \gamma$,
\begin{equation}\label{o1}
\begin{split}
& X_i(=0) = \sin(\alpha)\cos(\phi_i) \sigma_1 + \sin(\alpha)\sin(\phi_i) \sigma_2 + \cos(\alpha) \sigma_3, \\&
X_i(=1) = \sin(\beta)\cos(\phi_i+\gamma) \sigma_1 + \sin(\beta)\sin(\phi_i+\gamma) \sigma_2 \\& \hspace{1.7cm} + \cos(\beta) \sigma_3,
\end{split}
\end{equation} where $\sigma_i (i \in \{1,2,3\})$  are the Pauli matrices. Taking the values of $t \in [0,\frac{\pi}{4}]$ at small intervals, the QM expression of left-hand side of (\ref{rn}) for GHZ-like states (\ref{ghz}) is numerically maximized with respect to these parameters ($\phi_i, \alpha, \beta, \gamma$), as shown in Fig.2.

\begin{figure}[http]
\centering
\includegraphics[height=6cm,width=9cm]{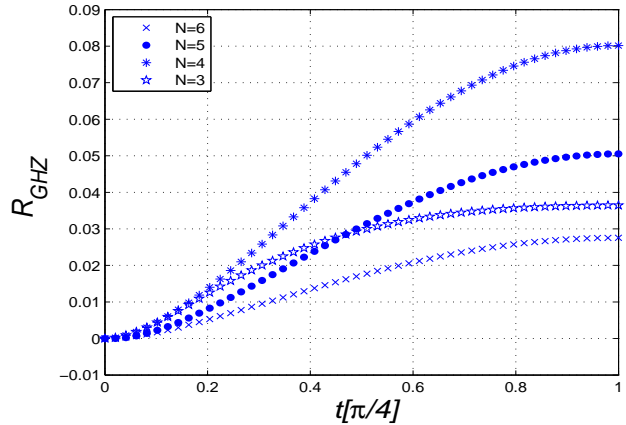}
\caption{QM violation of (\ref{rn}) for GHZ-like states (\ref{ghz}).}
\end{figure}
Remarkably, for any non-zero value of $t$, we get a QM violation. In other words, all the GHZ-like states (\ref{ghz}) possess genuine multipartite nonlocality in terms of violating first event broadcasting correlations. The analytical proof for the
tripartite case is given in the Appendix. These results signify the importance of the inequality (\ref{rn}).

\section{Implications and discussion} In this work, we were able to reveal a rich structure of different correlations in tripartite scenario and show the efficiency of QM versus local theories augmented with signaling. Apart from further studies on this kind of correlations our work opens an intriguing area of research as it hints that in scenarios with more parties the structure of correlations may be even more interesting.

Let us discuss one aspect of QM violation of inequalities satisfied by broadcasting correlations in terms of {\it anonymous quantum nonlocality} \cite{aqn}. This notion is based on the multipartite nonlocal correlations which can be reproduced by bilocal correlations with respect to all bipartitions. From an operational perspective, given that an outsider has access to the multipartite probability distributions, it is impossible to tell which subset of parties collaborated, even though the probability distribution is bilocal.  Obviously, {\it anonymous quantum nonlocality} can also be defined with respect to broadcasting. In the tripartite case, the correlations satisfying
\begin{equation}\label{aqn}
\begin{split}
& P(a,b,c|X,Y,Z)\\& = \sum_{\lambda_1} q_{\lambda_1}  P_{\lambda_1}(a|X) P^{X \rightarrow Y}_{\lambda_1}(b|Y,X,a) P^{X \rightarrow Z}_{\lambda_1}(c|Z,X,a) \\& = \sum_{\lambda_2} q_{\lambda_2} P_{\lambda_2}(b|Y) P^{Y \rightarrow X}_{\lambda_2}(a|X,Y,b) P^{Y \rightarrow Z}_{\lambda_2}(c|Z,Y,b) \\& = \sum_{\lambda_3} q_{\lambda_3} P_{\lambda_3}(c|Z) P^{Z \rightarrow Y}_{\lambda_3}(b|Y,Z,c) P^{Z \rightarrow X}_{\lambda_3}(a|X,Z,c) \\&
\neq  \sum_{\lambda} q_{\lambda} P_{\lambda}(a|X) P_{\lambda}(b|Y)  P_{\lambda}(c|Z)
\end{split}
\end{equation} are also anonymous in the sense that they can be reproduced by any kind of broadcasting but by local theory. Similarly, they can be defined in $N$-partite system with $k$ broadcasting parties.
To provide an example of quantum correlations satisfying (\ref{aqn}), we probe the same GHZ correlation considered in \cite{aqn}. The joint probability of all the measurements, denoted as $P(\vec{a}|\vec{X})$ for the GHZ state when the local measurements are $X_i(=0)=\sigma_1,X_i(=1)=\sigma_2$, is given by
\begin{equation}\label{ghzc}
\begin{split}
P(\vec{a}|\vec{X}) = \frac{1}{2^N} \left[1 + \cos \left(\frac{\pi }{2}\sum^N_{i=1} X_i \right) \prod^N_{i=1} a_i \right].
\end{split}
\end{equation} This correlation is nonlocal and has been studied extensively \cite{Mermin,ghzc}. In the $(N-2)$ event broadcasting scenario, irrespective of which $(N-2)$ parties are measured first, the probability distribution for all broadcasting parties satisfies $P(a^X_i+)=P(a^X_i-)=\frac{1}{2}$. Furthermore, the probability distribution for the last two parties is
\begin{equation}\begin{split}
& P(a_{N-1}^0-,a_{N-1}^1+,a_N^{y}=(l)(-)^{\lfloor{\frac{x}{2}}\rfloor+y+1},a_N^{y\oplus1}=(l)(-)^{\lceil{\frac{x}{2}}\rceil+y+1})\\
& = P(a_{N-1}^0+,a_{N-1}^1+,a_N^{y}=(l)(-)^{\lfloor{\frac{x}{2}}\rfloor+y},a_N^{y\oplus1}=(l)(-)^{\lceil{\frac{x}{2}}\rceil+y+1}) \\
& = P(a_{N-1}^0+,a_{N-1}^1-,a_N^{y}=(l)(-)^{\lfloor{\frac{x}{2}}\rfloor+y},a_N^{y\oplus1}=(l)(-)^{\lceil{\frac{x}{2}}\rceil+y}) \\
& = P(a_{N-1}^0-,a_{N-1}^1-,a_N^{y}=(l)(-)^{\lfloor{\frac{x}{2}}\rfloor+y+1},a_N^{y\oplus1}=(l)(-)^{\lceil{\frac{x}{2}}\rceil+y}) \\& = \frac{1}{4},
\end{split}\end{equation} where $x=modulo_4(\sum^{N-2}_{i=1}X_i),y=modulo_2(x),l=\prod^{N-2}_{i=1}a_i$. One can check that such $(N-2)$ event broadcasting reproduces the GHZ correlation (\ref{ghzc}).

In future, implications of QM violation of broadcasting in quantum information and communication can be investigated in more detail. \\

\section*{ Acknowledgements} We thank Marek \.{Z}ukowski and Piotr Mironowicz for helpful comments. This work is supported by the IDSMM programme at University of Gda\'{n}sk, the FNP programme TEAM, ERC grant QOLAPS and NCN grant 2013/08/M/ST2/00626.

\section*{Appendix}
Here, we show that all tripartite GHZ-like states violate (\ref{r}). Considering $\alpha=\beta=\frac{\pi}{2}$ and $\chi=\sum^N_{i=1}\phi_i$ in (\ref{o1}), the QM expression is obtained as
\begin{equation*}\begin{split}
& R_{3} = \frac{\sin(2t)}{8} [ \cos(\chi) - 3 \cos(\chi+\gamma)- 3\cos(\chi+2 \gamma)] - \frac{5}{8} \\
& = \sin(2t)0.6614-0.625 ~~(\text{taking } \chi =1.3807 , \gamma = 1.0472).
\end{split}\end{equation*}
For the QM violation of $R_3$, $t > \frac{1}{2}\sin^{-1}(\frac{0.625}{0.6614}) \approx 0.6187.$ While the maximum violation is 0.0364 when $t=\frac{\pi}{4}.$

To show the QM violation for lower $t$, we consider $\phi_i = 0 \forall i$ in (\ref{o1}), and
further, choose $ \cos(a) = t - 1$,
\begin{equation*}\begin{split}
& \cos(b) = \frac{(2t-t^2) \cos(2t)} {2t\cos(2t) + 4\sin^2(t) + (2t-t^2+2\sqrt{2t-t^2})\sin (2t)}.
\end{split}\end{equation*} Then the expression for $R_3$ given by
 \begin{equation*}
\begin{split}
& R_3 = \frac{\sin^2t}{8} [3(\cos(b)-1)(t-2)^2 + 3(\cos(b)+1)^2(t-2)\\ & - (t-2)^3]
 - \frac{\cos^2t}{8} [3t^2 (\cos(b)+1) + 3t(\cos(b)-1)^2 - t^3] \\
& - \frac{\sin(2t)}{8} [(2t-t^2)^{\frac{3}{2}} - 3\sin(b)(2t-t^2) - 3\sin^2(b) \sqrt{2t-t^2}]
\end{split}
\end{equation*} can be evaluated. It can be seen that for $t \in (0,0.66]$ this quantity is positive. Thus by taking two different ranges of values, we show that, for all $t\in (0,\frac{\pi}{4}]$, (\ref{r}) is violated.


\begin{thebibliography}{99}
\bibitem{cf} The correlation function is given by,
\begin{equation*} \begin{split}
& \langle abc \rangle = P(a+,b+,c+) -  P(a-,b+,c+) - P(a+,b-,c+) \\& - P(a+,b+,c-) + P(a-,b-,c+) + P(a-,b+,c-) \\& + P(a+,b-,c-) - P(a-,b-,c-).
\end{split}\end{equation*}
\end{thebibliography}
\end{document}